\documentclass[aps,pra,floatfix,twocolumn,superscriptaddress]{revtex4-1}
\usepackage{graphicx}
\usepackage{float}
\usepackage{subfigure}
\usepackage[english]{babel}
\usepackage{amsmath}
\usepackage{amssymb}
\usepackage{tensor}
\usepackage{color}

\newcommand{\be}{\begin{eqnarray}}
\newcommand{\ee}{\end{eqnarray}}
\newcommand{\p}{\partial}

\def\ep#1{\langle #1 \rangle}

\begin{document}

\title{Effects of heterostrain and lattice relaxation on optical conductivity of twisted bilayer graphene}

\author{ Zhen-Bing Dai}
\affiliation{College of Physics, Sichuan University, Chengdu, Sichuan 610064, China}
\affiliation{Department of Physics, Sichuan Normal University, Chengdu, Sichuan 610066, China}
\author{Yan He}
\affiliation{College of Physics, Sichuan University, Chengdu, Sichuan 610064, China}
\author{Zhiqiang Li}
\thanks{Electronic address: zhiqiangli@scu.edu.cn}
\affiliation{College of Physics, Sichuan University, Chengdu, Sichuan 610064, China}


\begin{abstract}
We present a theoretical study of the effects of heterostrain and lattice relaxation on the optical conductivity of twisted bilayer graphene near the magic angle, based on the band structures obtained from a continuum model. We find that heterostrain, lattice relaxation and their combination give rise to very distinctive spectroscopic features in the optical conductivity, which can be used to probe and distinguish these effects. From the spectrum at various Fermi energies, important features in the strain- and relaxation-modified band structure such as the bandgap, bandwidth and van Hove singularities can be directly measured. The peak associated with the transition between the flat bands in the optical conductivity are highly sensitive to the direction of the strain, which can provide direct information on the strain-modified flat bands.
\end{abstract}

\maketitle
\maketitle

\section{Introduction}

Recent experimental studies have discovered many new correlated electronic phases~\cite{CaoY1801,CaoY1802,Yankowitz19,Sharpe19,Lu19,Serlin20, XieY19,Jiang19,Kerelsky19} in twisted bilayer graphene (TBG) associated with the formation of flat electronic bands near some magic angles~\cite{MacDonald11,Morell10,Santos12}, which has generated great interest in this system. The flat bands and intriguing electronic phases created by interlayer coupling in the moir\'e superlattices exhibit critical dependence on small lattice deformations. Scanning tunneling experiments~\cite{Kerelsky19} and theoretical calculations~\cite{Huder18,Zhen19} have shown that strains with opposite sign in the two layers (heterostrain) can significantly modify the band structure and the observed electronic phases, through strain-induced effective gauge field and changes in stacking and interlayer tunneling. Extrinsic strain is ubiquitous in TBG, which is induced during the fabrication process and by the substrate. Moreover, the competition between interlayer van der Waals interaction and intralayer lattice distortion can cause significant atomic scale lattice relaxations in TBG involving local lattice rotations with localized strain to favor interlayer commensurability~\cite{ZhangK18,Carr18,Nam17,Yoo19,Alden13,DaiS16} , which induces substantial changes in lattice symmetry and electronic band structure. Therefore, it is of great importance to explore the effects of strain and lattice relaxation in TBG.

Optical experiment is a powerful technique to investigate the band structures, many body interactions and various electronic excitations in graphene-based systems~\cite{Basov11}. The optical conductivity of TBG has been studied theoretically and experimentally in several previous works~\cite{Tabert13,Stauber13,Moon14,Calderon20,Novelli20,Wen21,Nguyen17,Havener14,YuK19}. Earlier theoretical calculations ~\cite{Tabert13,Stauber13,Moon14,Moon13} have been done for twist angles $\theta>=1.47^\circ$. More recently, the optical conductivity near the magic angle has been calculated taking into account correlation effects ~\cite{Calderon20} and interlayer tunneling asymmetry ~\cite{Novelli20,Wen21}. So far, the combined effects of heterostrain and lattice relaxation on the optical properties of TBG are yet to be explored.

In this work, we show that the band structure of TBG at the magic angle is strongly affected by strain and lattice relaxation, which leads to very distinctive spectroscopic features in the optical conductivity, including the energy and lineshape of the peaks in the spectrum. Meanwhile, the bandgap, bandwidth and van Hove singularities can be directly measured from the spectrum by changing the Fermi energy. The peak associated with the transition between the flat bands are highly sensitive to the strain, which can provide direct information on the strain-modified flat bands. Therefore, the spectroscopic features in the optical conductivity can be used to probe and distinguish the effects of strain and lattice relaxation.

This paper is organized as follows: Section \ref{sec-bg} summarizes the effective continuum model taking into account the effects of strain and lattice relaxation, and describes the procedure used to calculate the optical conductivity. Section \ref{sec-results} presents our main findings on the band structure, DOS and optical conductivity. A summary and conclusions are reported in Section \ref{sec-conclusion}.

\section{Theoretical background}\label{sec-bg}

To investigate the effect of geometric deformation on the properties of TBG, we at first define the primitive lattice vectors of the initial monolayer graphene (MLG) before deformation as $A_1=a(1,0)$ and $A_2=a(1/2,\sqrt3/2)$, where $a\approx0.246$ nm is the lattice constant of MLG. The corresponding reciprocal lattice vectors are $G_1=\frac{2\pi}{a}(1,-1/ \sqrt3)$ and $G_2=\frac{2\pi}{a}(0,2/ \sqrt3)$. The low energy electronic structure of TBG
is calculated from an effective continuum model~\cite{Santos07,Santos12,Moon13,MacDonald11,Kindermann11}, taking into account the effects of heterostrain~\cite{Zhen19}. For one particular valley, the effective Hamiltonian of the continuum model is written in the basis of $(A_1, B_1, A_2, B_2)$ as
\be
H=\left(\begin{array}{cc}
              H_1 & U^{\dag}\\
              U & H_2
              \end{array}\right),\label{Ham}
\ee
where $A_l$ and $B_l$ are the two sublattices of the layer $l$ with $l=1,2$ representing two layers. Here, $H_l$ is the intralayer Hamiltonian given by the two-dimensional Weyl equation~\cite{Nam17,Zhen19}
\be
H_l=-\hbar v_F[(\mathbb{I}+\mathcal{E}_l^T)(\mathbf{k}-\mathbf{D}_l)\cdot(\sigma_x,\sigma_y),
\label{Ham12}
\ee
where $v_F$ is the Fermi velocity of MLG, $\sigma_x$ and $\sigma_y$ are Pauli matrices. We take $\hbar v_F/a\approx2.1354$ eV~\cite{Moon13,Koshino18}. The relative deformation matrix $\mathcal{E}_l$ includes both strain and rotation in general. $\mathbf{D}_l$ denotes the positions of the two Dirac fermions in momentum space under deformation, which is given by the previous studies~\cite{Nam17,Zhen19}. The effective interlayer coupling $U$ is given by ~\cite{Koshino18}
\be
U &&=\left(\begin{array}{cc}
              U_{A_2A_1} & U_{A_2B_1}\\
              U_{B_2A_1} & U_{B_2B_1}
              \end{array}\right)\nonumber\\
&&=\left(\begin{array}{cc}
              u & u'\\
              u' & u
              \end{array}\right)
+\left(\begin{array}{cc}
              u & u' \omega^{-\xi}\\
              u'\omega^{\xi} & u
              \end{array}\right)e^{i\xi \mathbf{G}_1^M\cdot \mathbf{r}}\nonumber\\
&&+\left(\begin{array}{cc}
              u & u' \omega^{\xi}\\
              u' \omega^{-\xi} & u
              \end{array}\right)e^{i\xi (\mathbf{G}_1^M+\mathbf{G}_2^M)\cdot \mathbf{r}},
              \label{coupling}
\ee
where $\mathbf{G}_l^M =\mathcal{E}^T \mathbf{G}_l(l=1,2)$ represents the reciprocal lattice vectors for the moire pattern, $\xi=\pm1$ is the valley index, $u$ and $u'$ describe respectively the amplitudes of diagonal and off-diagonal terms in the sublattice space, and $\omega=e^{2\pi i/3}$. For unrelaxed TBG, $u=u'=103$ meV~\cite{Moon13}. After atomic scale lattice relaxation, the band strcuture and physical properties in TBG were reported to have been significantly modified ~\cite{Nam17,Gargiulo17,Yoo19}. The effect of lattice corrugation (relaxation in the out-of-plane direction) in TBG can be included by choosing $u$=0.0797 eV and $u'$=0.0975 eV in the continuum model ~\cite{Koshino18}.

\begin{figure}[H]
\centerline{
\includegraphics[width=\columnwidth]{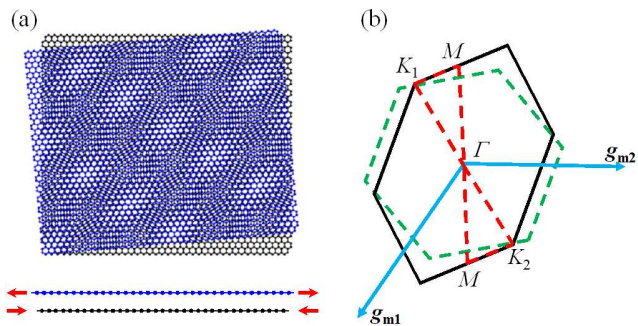}}
\caption{Sketch of TBG with uniform uniaxial strain along armchair direction and the resulting moire Brillouin zone. In (a), blue lines and black lines from the top views represent the top layer and bottom layer, respectively. Red arrows in the side views stand for the axial strain applied to each layer. The solid black lines and dashed green lines in (b) indicate the moire Brillouin zones of TBG with and without uniform strain, respectively. The blue arrows denote the reciprocal lattice vectors for the moire patterns. The dashed red lines show the path in the momentum space for the electronic structure.}
\label{fig-struc}
\end{figure}

Here, we focus on the effect of uniaxial heterostrain in TBG observed in scanning tunneling microscope (STM) experiments~\cite{XieY19,Jiang19,Kerelsky19}. With uniaxial strain, the deformation matrix $\mathcal{E}_l$ can be written as~\cite{Kerelsky19,Pellegrino10}
\be
\mathcal{E}_l &&=T(\theta)+S(\epsilon_l,\phi)\nonumber\\
&&=\epsilon_l \left(\begin{array}{cc}
              -\cos^2\phi+\nu \sin^2\phi & (1+\nu)\cos\phi \sin\phi\\
              (1+\nu)\cos\phi \sin\phi & -\sin^2\phi+\nu \cos^2\phi
              \end{array}\right)\nonumber\\
&&+\left(\begin{array}{cc}
              \cos\theta & -\sin\theta\\
              \sin\theta & \cos\theta
              \end{array}\right),
\label{epsilon}
\ee
where $T(\theta)$ represents rotation with angle $\theta$, the strain tensor $S(\epsilon_l,\phi)$ is characterized by the strain magnitude $\epsilon_l$ and the strain direction $\phi$. In this paper, we assume that $\mathcal{E}_1=-\mathcal{E}_2=\mathcal{E}/2$, namely the two layers are strained and rotated oppositely with the same magnitude. When a uniaxial heterostrain is applied to TBG, the moire superlattice in real space will be distorted from a regular triangular lattice as schematically shown in Fig. \ref{fig-struc}(a). Consequently, the resulting Brillouin zone is a distorted hexagon as shown in Fig. \ref{fig-struc}(b).

To study the optical conductivity, we calculate the eigenenergies and eigenstates in k-space by numerically diagonalizing the Hamiltonian within a limited wave space~\cite{Moon13}. Within the linear response theory, the Kubo formula for optical conductivity is given by~\cite{Stauber13}
\be
\sigma_{xx}(w)&&=\frac{e^2\hbar}{i (2\pi)^2}\int dk_x dk_y
\sum_{m,n}\frac{f(E_m)-f(E_n)}{E_m-E_n}\nonumber\\
&&\times\frac{|\ep{u_m|v_x|u_n}|^2}{\hbar\omega+i\eta+E_m-E_n},
\label{Eq-sig}
\ee
where $f(E_m)$ is the Fermi-Dirac distribution function $f(E_m)=1/(1+e^{(E_m-\varepsilon_F)/(k_B T)})$ with $E_m(E_n)$ and $|u_m\rangle(|u_n\rangle)$ representing the eigenenergies and eigenstates of TBG, respectively. The velocity operator is related to the Hamiltonian through the relation $v_x=\frac{1}{\hbar} \frac{\p H}{\p k_x}$, and $\eta$ denotes a phenomenological damping parameter.

\begin{figure*}
\centering
\subfigure
{\includegraphics[width=0.49\textwidth]{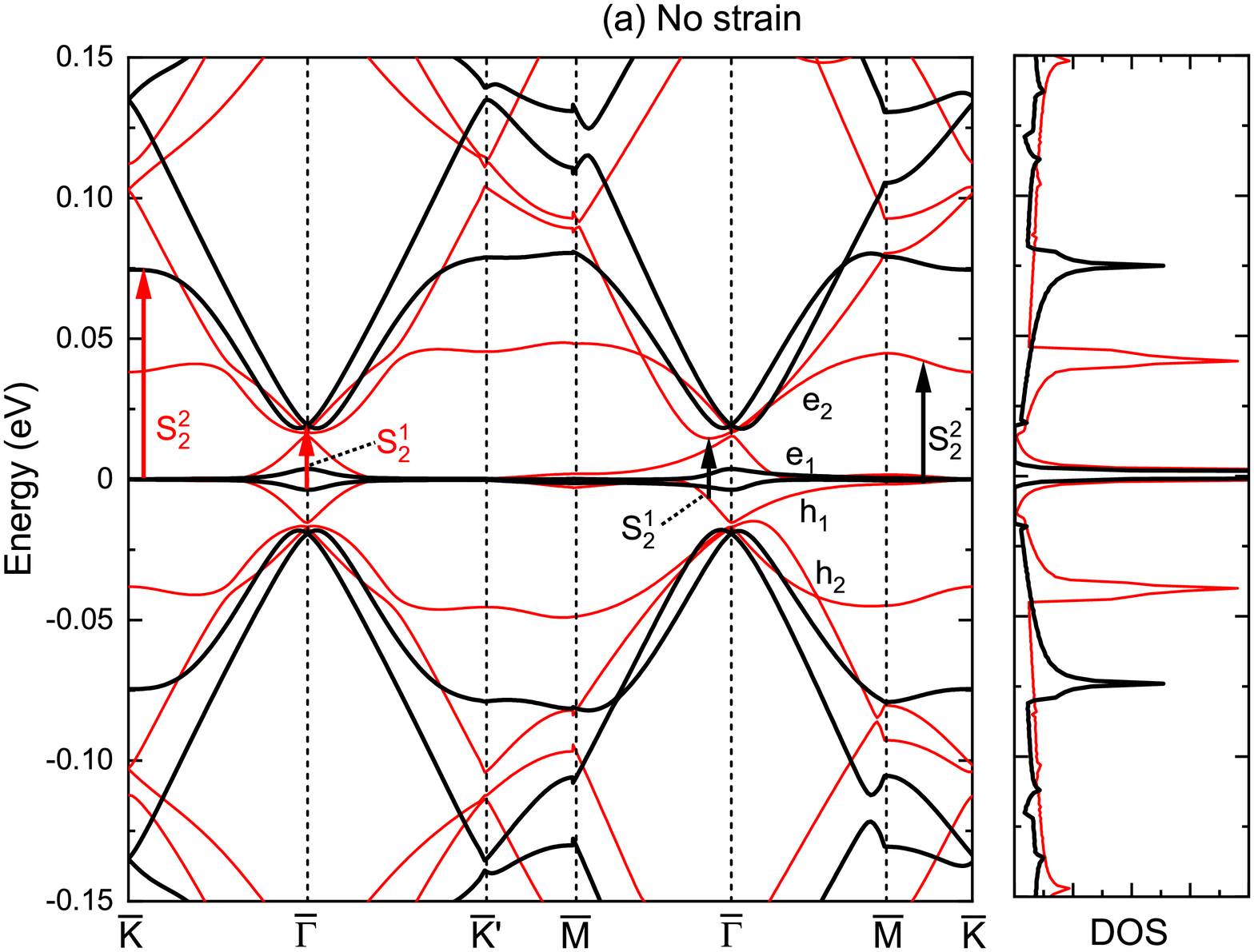}}
\subfigure
{\includegraphics[width=0.49\textwidth]{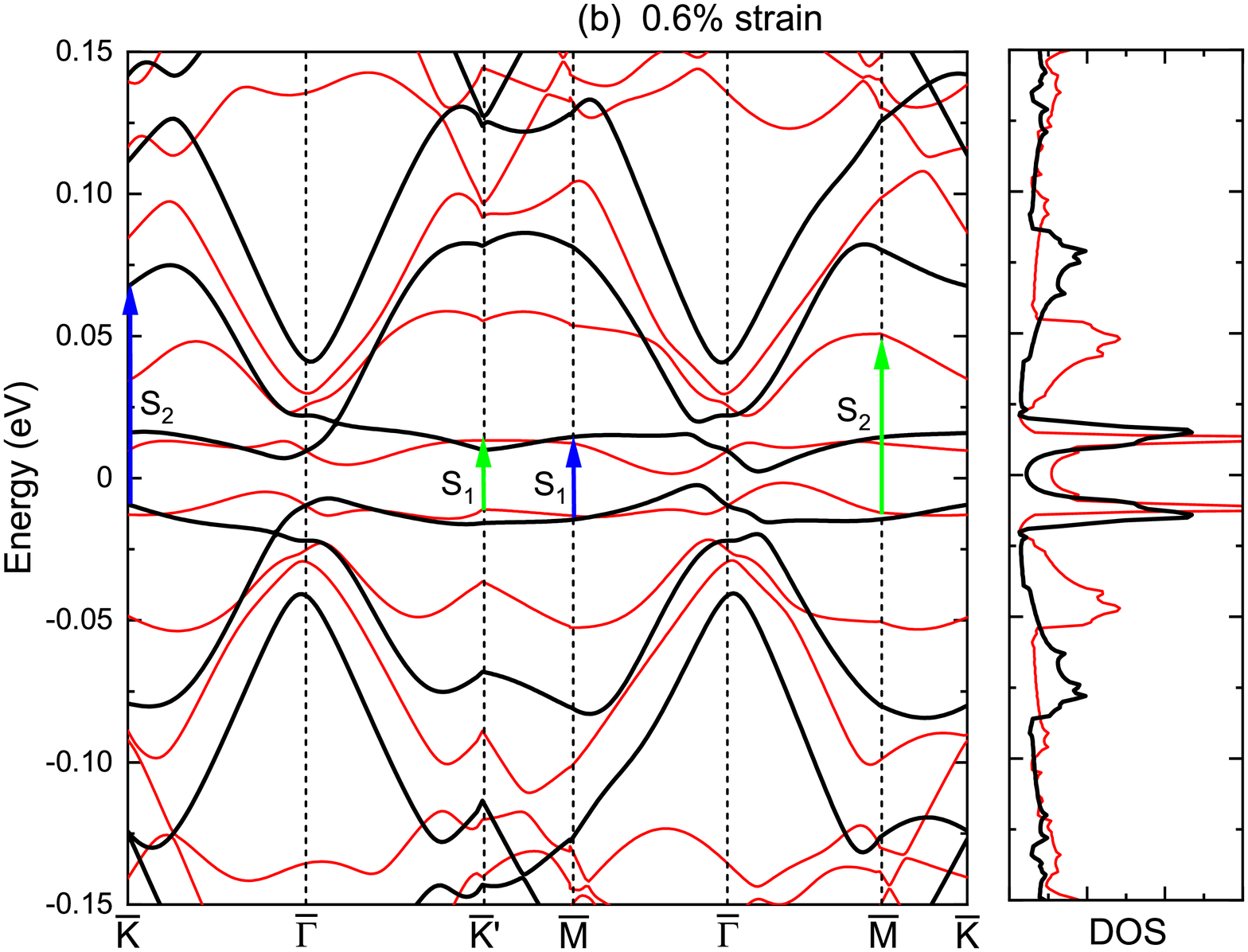}}
\caption{Band structure and DOS of TBG without/with the uniform heterostrain and relaxation at twist angle $\theta=1.05^\circ$. (a) and (b) show the influence of uniform heterostrain on the band structure (left) and DOS (right), in which the red lines and black lines correspond respectively to the cases without and with relaxation.
The heterostrain, $\epsilon=0.6\%$, applied to the graphene layers along a particular direction $\phi=30^\circ$.
The interband transitions between the saddle point of the lowest band ($h_1$ and $e_1$) and the band edge of the
second band ($h_2$ and $e_2$), which are marked by the colored arrows $S_2^1 (S_2^2)$ and $S_2$. And $S_1$ represents the transition between two split flat subbands near CNP.}
\label{fig-band}
\end{figure*}

\section{Results and discussions}\label{sec-results}
\subsection{Band structure}

The effects of heterostrain and lattice relaxation on the band structure of TBG at the magic angle $\theta=1.05^\circ$ are shown in Fig. \ref{fig-band}. In the absence of strain, the non-relaxed TBG exhibits flat bands with vanishing band velocity near the charge neutral point (CNP), leading to a strong peak in DOS near CNP. However, in the relaxed TBG, a band gap opens up between the flat bands near CNP and the excited bands in both the electron and hole sides. The bandwidth of the flat bands becomes even narrower and the band velocity is reduced compared to the nonrelaxed case. The excited bands and the associated van Hove singularities are moved to higher energies when the lattice relaxation is taken into account. These results are similar to previous studies~\cite{Koshino18}. In the presence of strain, the band structure is strongly affected. Comparing the non-relaxed cases with and without strain, the energy separation between the lowest conduction and valence bands is significantly enlarged in the former case~\cite{Zhen19}, which can be seen from the separation of two van Hove singularities near CNP. While these two bands are quite flat in most regions of the Brillouin zone, they remain connected by two Dirac crossings in each valley~\cite{Zhen19}. The high energy van Hove singularities are generally broadened in the strained case compared to the non-strained case.

The band structure and DOS of strained TBG are influenced by lattice relaxation. As shown in Fig. \ref{fig-band}(b), the van Hove singularities of the lowest conduction and valence bands in relaxed TBG are broader compared to the non-relaxed TBG. Moreover, the excited bands shift to higher energies under the influence of lattice relaxation. As to the DOS, such a shifting is the most pronounced for van Hove singularities associated with the first excited bands in both electron and hole sides. We note that heterostrain and lattice relaxation induce strong features of the DOS in different energy regions. The most prominent feature due to strain is the large splitting of two van Hove singularities near CNP, whereas the strongest feature due to lattice relaxation is the blue-shift of van Hove singularities associated with the first excited bands.

\begin{figure*}
\centering
\subfigure
{\includegraphics[width=1.0\columnwidth]{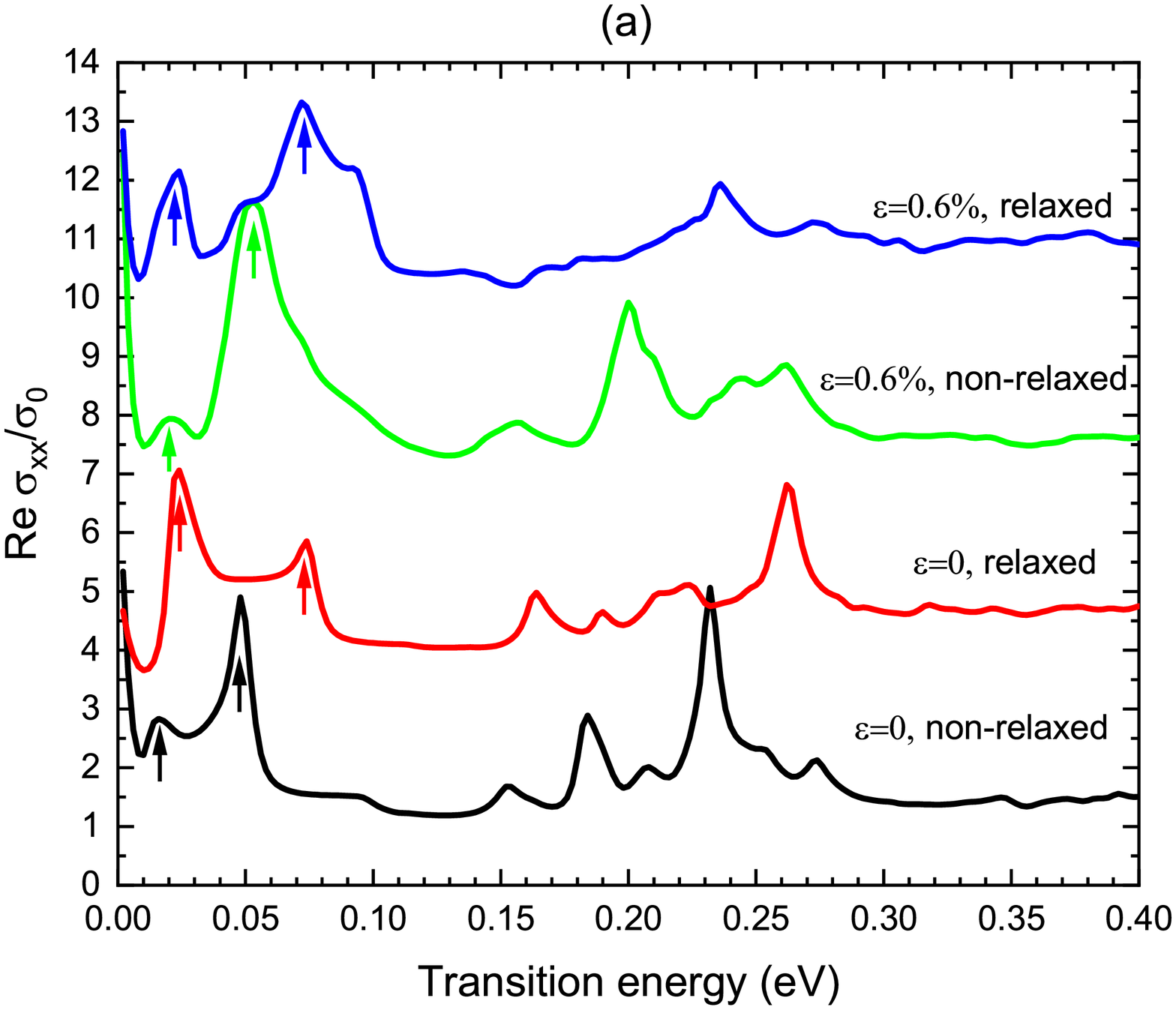}}
\subfigure
{\includegraphics[width=1.0\columnwidth]{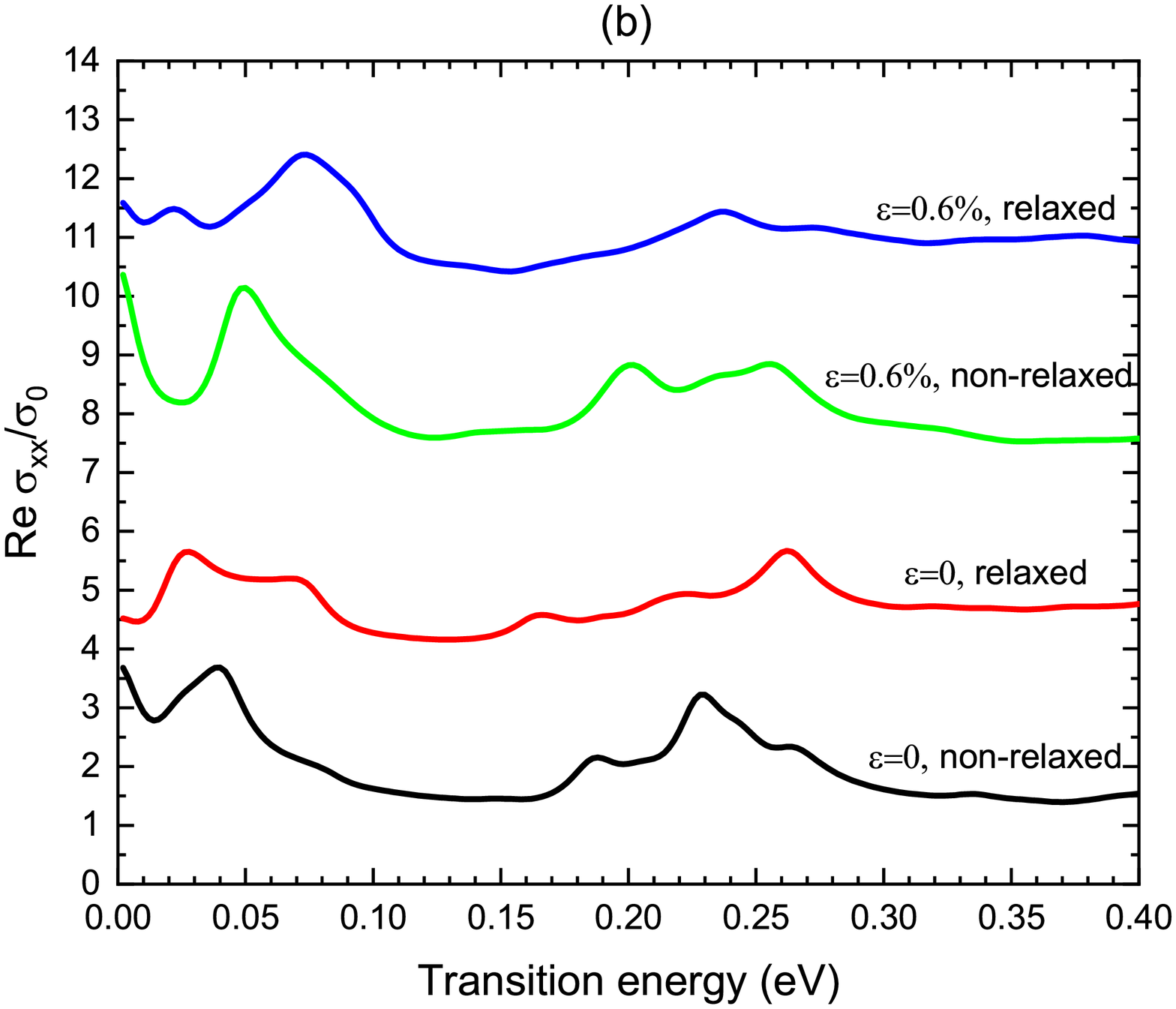}}
\caption{Real part of optical conductivity of TBG for the four typical cases shown in Fig. \ref{fig-band}.
The phenomenological damping rate $\eta$ is chosen to be 3 meV in (a) and 10 meV in (b). In the realistic situation, the value of $\eta$ depends usually on the electronic mobility of the sample. The arrows indicate the transitions
shown in the band structure Fig. \ref{fig-band}. The spectra are offset in $\sigma_{xx}$ by $3\sigma_0$ apiece for clarity. The temperature is set to be $T=10$ K in the numerical calculation. $\sigma_0=e^2/4\hbar$.}
\label{fig-oc-eta}
\end{figure*}

\subsection{Optical conductivity}

In order to reveal the influence of strain and lattice relaxation, we compare the optical conductivity of TBG for non-relaxed TBG without strain, non-relaxed TBG with strain, strained TBG without relaxation, and strained TBG with relaxation. The optical conductivity are scaled by $\sigma_0=e^2/4\hbar$, which is the background conductivity of monolayer graphene. Spectra calculated using two different damping parameters $\eta$ are shown in order to illustrate the effect of electronic mobility of TBG samples, with $\eta=3$ meV ($\eta=10$ meV ~\cite{LuoY20}) corresponding to a high (low) mobility sample. The optical conductivity spectrum exhibits two groups of peaks in the energy ranges of $0-100$ meV and $150-300$ meV. The second group of peaks are relatively weak and broad especially in low mobility samples ($\eta=10$ meV), so the change of these features due to strain and lattice relaxation will be difficult to detect in realistic samples. On the other hand, the first group of peaks are strong and show dramatic changes under the influence of strain and lattice relaxation in all spectra, which are easier to detect in optical measurements regardless of the mobility of samples. Hence, we will focus on peaks below 100 meV in our discussion.

In the non-relaxed TBG without strain ~\cite{Moon13,Moon14}, the optical conductivity exhibits two prominent peaks $S_2^1$  and $S_2^1$ below 100 meV arising from the interband transitions between flat bands and the first excited bands ($h_1 \rightarrow e_2, h_2 \rightarrow e_1$) as shown in Fig. \ref{fig-band}. This assignment is obtained from calculating the contributions to $\sigma_{xx} (w)$ from transitions involving different pairs of bands. The contribution to Re $\sigma_{xx} (w)$ from $h_1 \rightarrow e_1$ transition at finite energy is a smooth tail without a peak.

The optical conductivity spectrum is significantly changed by lattice relaxation effects (red lines in Fig. \ref{fig-oc-eta}): the $S_2^1$ peak becomes much stronger and the $S_2^2$ peak shifts to higher energy compared to the nonrelaxed case. The shift of the $S_2^2$ peak is due to the blue-shift of van Hove singularities associated with the first excited bands under lattice relaxation. In non-strained low mobility samples (Fig. \ref{fig-oc-eta}(b)), the $S_2^1$ and $S_2^2$ peaks are broadened and merge into a single broad peak below 100 meV in the nonrelaxed case, and lattice relaxation splits these two peaks with a large energy separation. Such a large change is readily observable in optical measurements of realistic samples.

The application of strain leads to large modifications in the optical conductivity. In strained TBG without lattice relaxation (green lines in Fig. \ref{fig-oc-eta}), the interband transition $h_1 \rightarrow e_1$ gives rise to a small $S_1$ peak because strain splits these two bands with a finite energy separation in most regions of the Brillouin zone. Moreover, under the influence of strain several interband transitions including $h_2 \rightarrow e_1, h_1 \rightarrow e_2, h_3 \rightarrow e_1, h_1 \rightarrow e_3, h_2 \rightarrow e_2, h_3 \rightarrow e_2, h_2 \rightarrow e_3$ produce narrow peaks in overlapping energy ranges around 50 meV, giving rise to a very strong peak $S_m$ at higher energy compared to the $S_2^2$ peak in non-relaxed TBG without strain. Comparing non-relaxed low mobility samples (Fig. \ref{fig-oc-eta}(b)), the peak below 50 meV in the non-strained case shifts to higher energy and gains a high-energy shoulder with strain. The peaks between 150 and 300 meV in non-relaxed cases are also strongly modified by strain.

Further inclusion of lattice relaxation effects in strained TBG makes the $S_1$ peak substantially stronger (blue lines in Fig. \ref{fig-oc-eta}), because the separated $h_1$ and $e_1$ bands become almost parallel in many regions of the of the Brillouin zone under lattice relaxation, leading to very strong joint DOS at the energy of the $S_1$ peak. Furthermore, the $S_m$ peak in non-relaxed strained TBG shifts to higher energy by lattice relaxation due to the blue-shift of van Hove singularities associated with the excited bands. In low mobility samples with strain (Fig. \ref{fig-oc-eta}(b)), the non-relaxed case exhibits a single peak with a high-energy shoulder below 100 meV, whereas there are two well-defined peaks in this energy range in the relaxed case. Such a large contrast will be easy to observe experimentally. The totality of the results in Fig. \ref{fig-oc-eta} demonstrates that the dramatic changes of peaks in optical conductivity can serve as spectroscopic signatures to probe and distinguish the effects of strain and lattice relaxation and explore the resulting band structure.

\begin{figure}[H]
\centering
\subfigure
{\includegraphics[width=1.0\columnwidth]{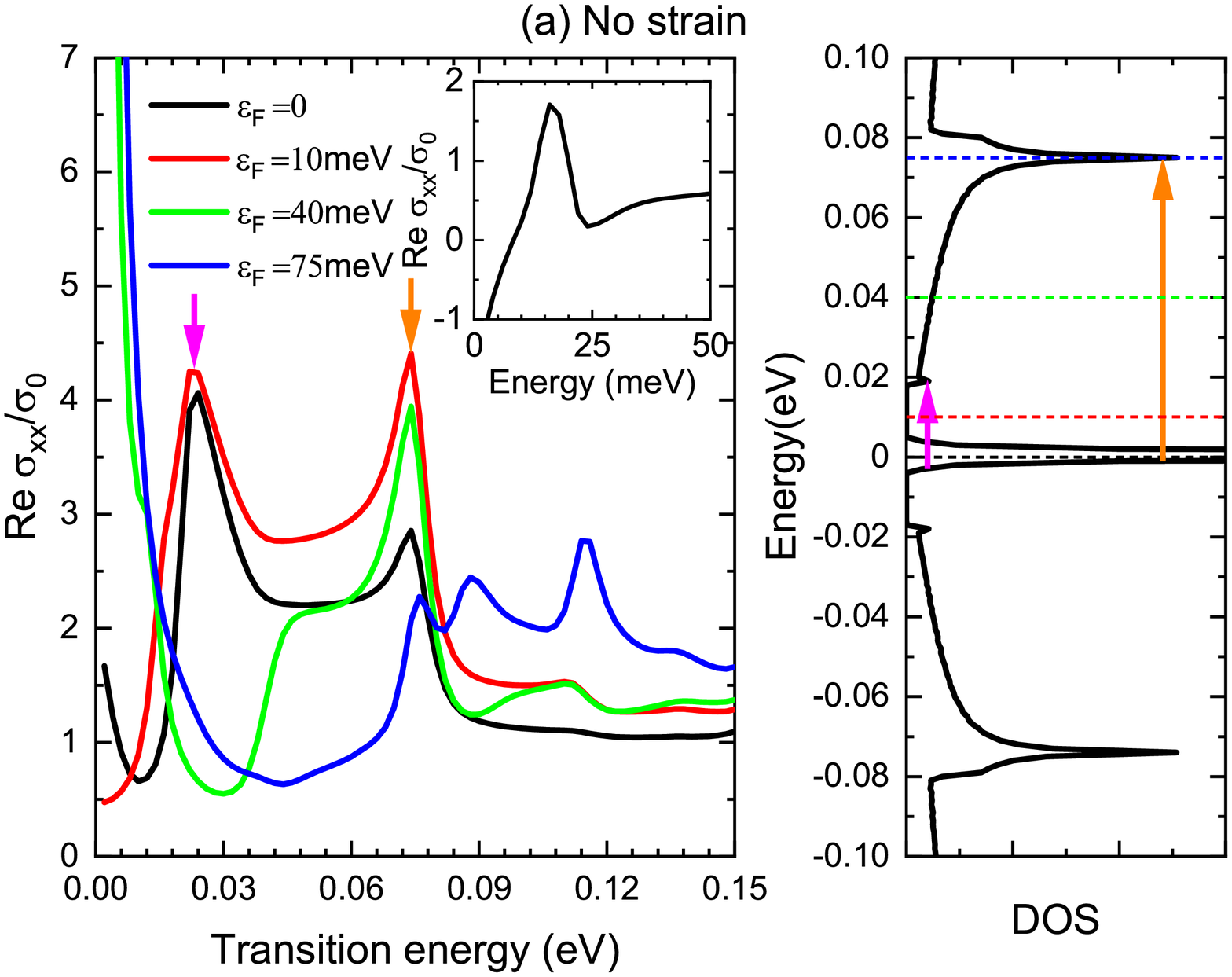}}
\subfigure
{\includegraphics[width=1.0\columnwidth]{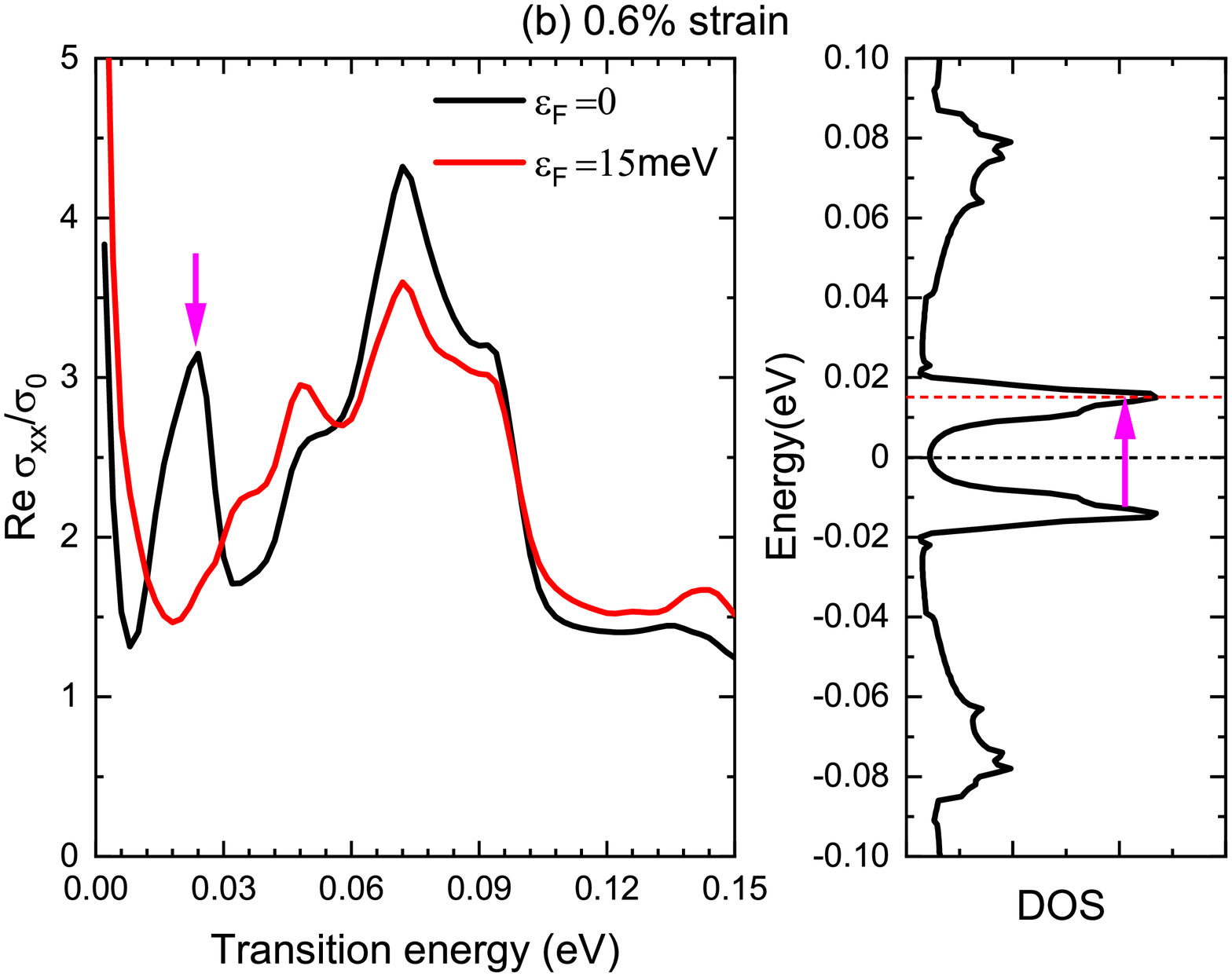}}
\caption{Real part of optical conductivity of relaxed TBG without strain (a) and with strain $\epsilon=0.6\%$ (b),
for various Fermi energies $\varepsilon_F$. The first significant van Hove singularity is at $E\simeq$75 meV in the
DOS in Fig. \ref{fig-band}(a) and at $E\simeq$15 meV in Fig. \ref{fig-band}(b), respectively. Various significant optical transitions are indicated by the colored arrows. The Fermi energies are marked as the dashed lines in the
DOS. The inset in (a) shows the relative spectrum of Re $\sigma_{xx}$ at $\varepsilon_F=10$ meV by subtracting Re $\sigma_{xx}$ at $\varepsilon_F=0$ as a function of transition energy. The phenomenological damping rate and the temperature are set to be $\eta=3$ meV and $T=10$ K, respectively.}
\label{fig-oc-EF}
\end{figure}

Important features in the band structure induced by strain and lattice relaxation can be explored from optical conductivities with various Fermi energies. Fig. \ref{fig-oc-EF}(a) shows the optical conductivity of relaxed TBG without strain with $\eta$=3 meV for different Fermi energies. When Fermi energy is moved into the bandgap between the flat bands and the neighboring bands, a new peak $S_\Delta$ appears below the $S_2^1$ peak compared to the spectrum with $\varepsilon_F=0$, which is clearly shown in the inset of Fig. \ref{fig-oc-EF}(a). The peak $S_\Delta$ arises from the $e_1 \rightarrow e_2$ transition and becomes active with $\varepsilon_F$ in the gap, therefore it provides a direct measure of the bandgap. Moreover, the energy separation between $S_\Delta$ and $S_2^1$ is a measure of the total bandwidth of the two flat bands. As $\varepsilon_F$ is further increased up to the next van Hove singularity around 75 meV, the $S_2^1$ peak disappears and then the $S_2^2$ peak weakens significantly due to Pauli blocking. Conversely, lowering the $\varepsilon_F$ for a highly doped sample can activate these transitions, which is a measure of the related van Hove singularity. For the strained relaxed TBG (Fig. \ref{fig-oc-EF}(b)), the lowest energy $S_1$ peak disappears as the Fermi energy is increased above the $e_1$ band, due to Pauli blocking. In this case, the $S_1$ peak provides information on the energy separation of the two flat bands induced by strain. The results in Fig. \ref{fig-oc-EF} demonstrates that important features in the band structure such as the bandgap, bandwidth and van Hove singularities can be directly measured from the optical conductivity spectrum by changing the Fermi energy.

\begin{figure*}
\centering
\subfigure
{\includegraphics[width=\textwidth]{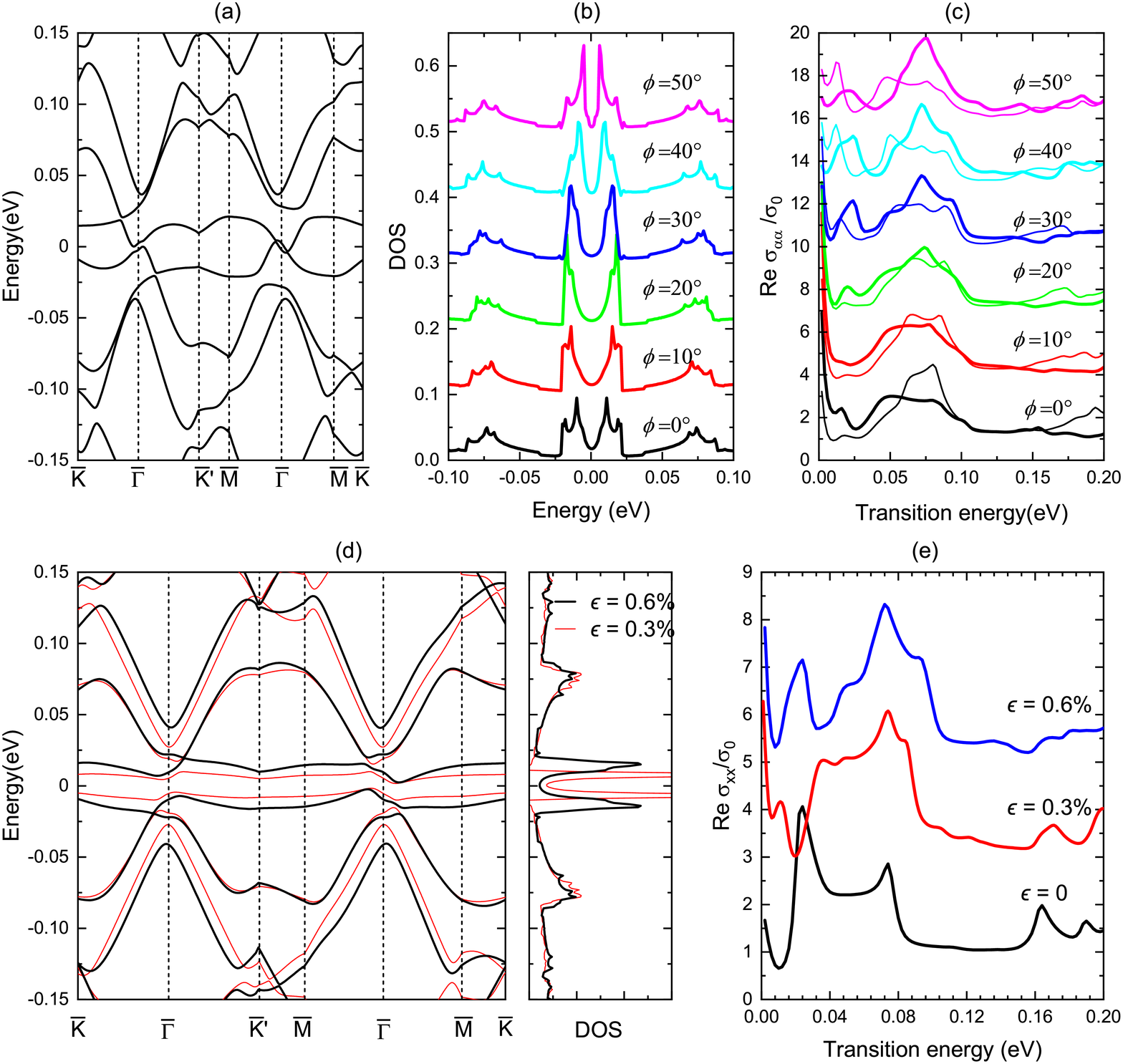}}
\caption{Influence of heterostrain on the band structure, DOS and optical conductivity. (a) Band structure of the relaxed TBG with strain $\epsilon=0.6\%, \phi=0^\circ$ at the magic angle. (b) DOS of relaxed TBG at $\theta=1.05^\circ$ with fixed strain $\epsilon=0.6\%$, and varying the direction of strain $\phi$. (c) Real part of optical conductivity $\sigma_{\alpha\alpha}$, where $\alpha=x$ (thick lines) or $y$(thin lines), of relaxed TBGs with the different directions of strain $\phi\in[0,60^\circ)$. The spectra are offset in $\sigma$ by $3\sigma_0$ apiece for clarity. (d) Band structure of relaxed TBG at $\theta=1.05^\circ, \phi=30^\circ$ with $\epsilon=0.3\%$ (red thin lines) and $\epsilon=0.6\%$ (black thick lines). (e) Real part of longitudinal optical conductivity of relaxed TBG with $\phi=30^\circ$ for the various strain magnitude. The first absorption peak associated with the transition $S_1$ shows dramatic changes under the influence of strain. The spectra are offset in $\sigma$ by $2\sigma_0$ apiece for clarity. The temperature and the phenomenological damping rate are set to be $T=10$ K and $\eta=3 $ meV, respectively.}
\label{fig-oc-angle}
\end{figure*}

Now we study the dependence of the DOS and optical conductivity on the direction of the strain $\phi$ in relaxed TBG. The DOS with fixed $\theta=1.05^\circ, \epsilon=0.6\%$ for several values of $\phi\in[0,60^\circ)$ are shown in Fig. \ref{fig-oc-angle}(b). The system is periodic for $\phi \rightarrow \phi + 60^\circ$ ~\cite{Zhen19}. As $\phi$ is varied, we observe interesting changes in the structure of van Hove singularities within $h_1$ and $e_1$ bands (between -20 and 20 meV) evolving from multiple DOS peaks to one prominent peak and then to multiple peaks in both bands, consistent with previous studies~\cite{Zhen19,Choi19,Kerelsky19}. The bandwidth of these two bands stays approximately constant. As shown in Fig. \ref{fig-oc-angle}(c), the Re $\sigma_{xx} (w)$ and Re $\sigma_{yy} (w)$ spectra are different for all values of $\phi$. This anisotropy is a result of the uniaxial strain considered in our calculation. Both Re $\sigma_{xx} (w)$ and Re $\sigma_{yy} (w)$ spectra are very sensitive to the strain direction. The peak $S_1 (h_1 \rightarrow e_1)$ around 20 meV in Re $\sigma_{xx} (w)$ and a peak around 10 meV in $\sigma_{yy} (w)$ become most prominent for $\phi$ near $30^\circ$ and $40^\circ$ and disappear for $\phi$ near $0^\circ$ and $10^\circ$. We find that the strength of these low energy peaks are related to the band structure and DOS. For $\phi$ around $30^\circ$ and $40^\circ$, the $h_1$ and $e_1$ bands are almost parallel in most regions of the Brillouin zone (Fig. \ref{fig-band}(b)), leading to one prominent DOS peak in their associated van Hove singularities, which gives rise to very pronounced peaks in Re $\sigma_{xx} (w)$ and Re $\sigma_{yy} (w)$ below 20 meV. For $\phi$ around $0^\circ- 10^\circ$, the $h_1$ and $e_1$ bands are no longer parallel (Fig. \ref{fig-oc-angle}(a)) leading to multiple DOS peaks in the associated van Hove singularities, so the low energy peaks in Re $\sigma_{xx} (w)$ and Re $\sigma_{yy} (w)$ disappear due to the lack of well-defined peaks in the joint DOS. Therefore, the low energy peaks in optical conductivity can provide information on the shape of strain-modified flat bands and the associated van Hove singularities.

The flat bands $h_1$ and $e_1$ are most sensitive to strain as shown in Fig. \ref{fig-oc-angle}(d), and their energy separation increases with strain. The other bands are less affected by strain. Fig. \ref{fig-oc-angle}(e) displays the evolution of Re $\sigma_{xx} (w)$ with strain. With increasing strain, the $S_1$ peak ($h_1 \rightarrow e_1$ transition) appears gradually and shifts to higher energy, because of the increasing energy separation between the $h_1$ and $e_1$ bands. Therefore, this peak can be used to estimate the magnitude of strain. The $S_2^1$ and $S_2^2$ peaks in the non-strained case shift to higher energy with strain and merge into a peak with shoulders on both sides.

In our work, a particular type of lattice relaxation, lattice corrugation in the out-of-plane direction, is included in our calculations by setting $u=0.0797$ eV and $u'=0.0975$ eV in the continuum model~\cite{Koshino18}. Previous theoretical studies ~\cite{Nam17} have shown that in-plane lattice relaxation modifies the band structure and DOS in a similar way to the changes induced by lattice corrugation shown in Fig. \ref{fig-band}(a), including the openning of a bandgap above and below the flat bands, the reduction of the bandwidth of the flat bands, and the blue-shift of the excited bands and the associated van Hove singularities. Therefore, we expect that the general trends in the changes of optical conductivity found in our studies will qualitatively be the same when the in-plane lattice relaxation is taken into account, which will be left for future work.

\section{Conclusion}\label{sec-conclusion}

In this paper, we systematically studied the effects of uniaxial heterostrain and lattice relaxation on the electronic structure and optical conductivity of TBG at the magic angle, which was calculated based on the band structure obtained from a widely used continuum model. We calculated the optical conductivity using different damping rates in order to facilitate comparisons with future optical measurements of TBG samples with different electronic mobilities. It was found that heterostrain, lattice relaxation and their combination lead to very distinctive spectroscopic features in the optical conductivity, including the energy and lineshape of the peaks in the spectrum, which can be used to probe and distinguish the effects of strain and lattice relaxation.

The heterostrain generically broadens the bandwidth of the nearly flat bands, which had been verified via energy separation of van Hove singularities in STM experiments. Moreover, different interband transitions can be activated or blocked at various Fermi energies, so important features in the band structure such as the bandgap, bandwidth and van Hove singularities can be directly measured from the spectrum by changing the Fermi energy. In the presence of heterostrain, we investigated the anisotropy of the conductivity spectra. Remarkably, we found that the absorption peaks associated with the transition between the flat bands are highly sensitive to the direction and magnitude of strain, which can provide direct information on the strain-modified flat bands. Therefore, our results can provide some insights into understanding the optical properties of TBG and offer a comparison of the optical spectrum of TBG in future experiment for detecting the imperceptible changes of the lattice structure.

In realistic TBG samples, the extrinsic strain and the degree of lattice relaxation generally change spatially. Spatial dependent measurements of the local optical conductivity using infrared microscopy can provide important insights into the local band structure and DOS modified by strain and lattice relaxation.

\begin{acknowledgments}

We are grateful to Guo-Yu Luo, Lu Wen, and Prof. Hao-Ran Chang for valuable discussions. This work was supported by the National Natural Science Foundation of China under Grant Nos. 11874271 and 11874272. We thank the High Performance Computing Center at Sichuan Normal University.
\end{acknowledgments}


\end{document}